\documentclass[graybox]{svmult}

\usepackage{type1cm}        
\usepackage{makeidx}         
\usepackage{graphicx}        
\usepackage{multicol}        
\usepackage[bottom]{footmisc}

\usepackage{newtxtext}       %
\usepackage{newtxmath}       

\makeindex             

\begin{document}

\title*{Dynamical systems on hypergraphs}
\author{Timoteo Carletti and Duccio Fanelli}
\institute{Timoteo Carletti \at naXys, Namur Institute for Complex Systems, University of Namur, rue de Bruxelles 61, B5000 Namur, Belgium, \email{timoteo.carletti@unamur.be}
\and Duccio Fanelli \at Universit\`a degli Studi di Firenze, Dipartimento 
di Fisica e Astronomia, CSDC and INFN, via G. Sansone 1, 50019 Sesto Fiorentino, Italy \email{duccio.fanelli@unifi.it}}
%
%
\maketitle

\abstract{We present a general framework that enables one to model high-order interaction among entangled dynamical systems, via hypergraphs. Several relevant processes can be ideally traced back 
to the proposed scheme. We shall here solely elaborate on the conditions that seed the spontaneous emergence of patterns, spatially heterogeneous solutions resulting from 
the many-body interaction between fundamental units. In particular we will focus, on two relevant settings. First, we will assume long-ranged mean field interactions between populations, and then turn to
considering diffusive-like couplings. Two applications are presented, respectively  to a generalised Volterra system and the Brusselator model.}


\section{Introduction}
\label{sec:intro}
The study of many body interactions has a long history in science and technology, and relevant results have been obtained under the assumption of regularity of the underlying substrates, where the dynamics eventually develops. When regularity gets lost, general results are scarce and simplifying assumptions, which implement dedicated approximations, need to be put forward. It is for instance customary to reduce the many body exchanges within a pool of simultaneously interacting entities to a vast collection of pairwise contacts, a working ansatz which drastically reduces the intimate complexity of the scrutinised dynamics. Governing dynamical systems are hence cast on top of networks~\cite{AlbertBarabasi,BLMCH} with diverse and variegated topologies: each node contains a replica of the original system,
 and the strength of interaction is set by the weight of the associated link.

Despite this crude approximation,  relevant results have been obtained which bear general interest~\cite{Newmanbook,Barabasibook,Latorabook}. At the same time many examples of systems exist for which the above assumption holds true just as a first order approximation~\cite{BGL,LRS2019}. To overcome this intrinsic limitation, the effect of aggregated structures of nodes, such 
as cliques, modules or communities~\cite{Newmanbook,fortunatohric2016} has been recently addressed in the literature. This implies analysing the cooperative interference within bunches of tightly connected nodes and assessing their role in shaping the ensuing dynamics, in the framerwok of a generalized picture which accounts for multiple pairwise exchanges. 

There are however several examples where the interactions among individuals, being them neurons~\cite{petri2014homological,LEFGHVD2016}, proteins~\cite{estradaJTB}, animals~\cite{Abrams1983,GBMSA} or  authors of scientific papers~\cite{PPV2017,CarlettiEtAl2020}, cannot be reduced to binary 
interactions. The group action is indeed the real driver of the dynamics. Starting from this observation, higher-order models have been developed so as to capture the many body interactions among individual units. We hereby focus on hypergraphs~\cite{berge1973graphs,estrada2005complex,GZCN}, versatile tools with a broad potential that is still being fully elucidated. Hypergraphs have been applied to different fields from social contagion model~\cite{de2019social,ATM2020}, to the modelling of random walks~\cite{CarlettiEtAl2020}, from the study of synchronisation~\cite{Krawiecki2014,MKJ2020,CarlettiJPC_2020} and diffusion~\cite{ATM2020}, to non-linear consensus \cite{neuhauser2020multibody}, via the emergence of Turing patterns~\cite{CarlettiJPC_2020}. It is also worth mentioning an alternative approach to high-order interactions which exploits the notion of simplicial complexes~\cite{DVVM,BC,PB}. Largely used in the past to tackle optimisation or algebraic problems, they have been recently invoked to address problems in epidemic spreading~\cite{BKS2016,IPBL} or synchronisation phenomena~\cite{LCB2020,GdPGLRCFLB2020,PhysRevLett.124.218301}. In this work we will however adopt the viewpoint of 
hypergraphs, to represent high-order interactions. 

Hypergraphs constitute indeed a very flexible paradigm. An arbitrary number of agents are allowed to interact: an hyperedge grouping all the involved agents  encodes for the many body interaction, thus extending conventional network models beyond the limit of binary contacts. A hypergraph can reproduce, in a proper limit, a simplicial complex and, in this respect, provides a more general tool for  addressing many body simultaneous interactions.

Based on the above, it can be claimed that many body interactions constitute a relevant and transversal research field that is still in its embryonic stage, in particular as concerns studies that relate to hypergraphs. Our contribution is positioned in this context and aims at systematising the study of   dynamical systems coupled via a hypergraph. For a sake of definitiveness, we will hereby consider the interactions to be mediated by the hyperedges, that is by the (hyper)adjacency matrix (see Section~\ref{sec:hyperg}), or by a diffusive-like process, that is implemented via a properly engineered Laplace matrix (see Section~\ref{sec:lap}). In both cases, we will be interested in the emergence of spatially heterogeneous solutions, i.e. coherent and extended patterns.

\section{Hypergraphs and high-order interactions.}
\label{sec:hyperg}

The aim of this section is to introduce the formalism of (hyper) adjacency matrix which enables us to account for the high-order interaction among 
several identical dynamical systems. We will then present a first study on the emergence of spatial heterogeneous solutions, i.e. patterns, for systems interacting via a hypergraph, by assuming that uncoupled individual 
units do converge to a (spatially) homogeneous stable solution.

\subsection{Hypergraphs}
\label{ssec:hyperg}
An hypergraph $\mathcal H(V,E)$ is defined by a set of nodes, $V=\{v_1,\dots,v_n\}$, and a set of $m$ hyperedges $E=\{E_1,\dots,E_m\}$, such that for all $\alpha=1,\dots,m$ : ${E_\alpha}\subset V$. If all hyperedges have size $2$ then the hypergraph reduces to a network. A simplicial complex is recovered if each hyperedge contains all its subsets.

One can encode the information on how the nodes are shared among hyperedges, by using the {\em incidence matrix of the hypergraph}~\footnote{We will adopt the convention of using roman indexes for nodes and greek ones for edges.}, $e_{i \alpha}$, namely
\begin{equation}
\label{eq:incid}
e_{i \alpha}=\begin{cases} 1 &\text{$v_i\in E_{\alpha}$}\\
0 & \text{otherwise}\, .
\end{cases}
\end{equation}

Given the latter, one can construct the $n\times n$ hypergraph adjacency matrix, 
\begin{equation}
\label{eq:hyperadj}
\mathbf{A}=\mathbf{e}\,\mathbf{e}^{\top}\, ,\quad A_{ij}=\sum_\alpha e_{i\alpha}e_{j\alpha}\, ,
\end{equation}
thus $A_{ij}$ represents the number of hyperedges containing both nodes $i$ and $j$. Let us observe that often in the literature the adjacency matrix is defined by imposing a null diagonal. In the following we will adopt a different notation by defining its diagonal to contain all $1$'s. This in turn amounts to assume the hypergraph to contain all the trivial hyperedges made of just a single node. Finally we define the $m\times m$ hyperedges matrix 
\begin{equation}
\label{eq:hyperC}
\mathbf{C}=\mathbf{e}^{\top}\mathbf{e}\, ,\quad C_{\alpha \beta}=\sum_ie_{i\alpha}e_{i\beta}\, ,
\end{equation}
$C_{\alpha \beta}$ counts the number of nodes in $E_{\alpha}\cap E_{\beta}$, hence $C_{\alpha\alpha}$ is the size of the hyperedge $E_{\alpha}$.

\subsection{High-order coupling}
\label{ssec:hhcoup}
Let us consider a $d$-dimensional dynamical system described by the ODE :
\begin{equation}
\label{eq:sys}
 \frac{d\mathbf{x}}{dt}(t)=\mathbf{f}(\mathbf{x}(t))\, ,
\end{equation}
where $\mathbf{x}(t)=(x_1(t),\dots,x_d(t))^\top$ denotes the state of the system at time $t$ and $\mathbf{f}$ is a generic nonlinear function which describes the rate of variation of $\mathbf{x}$. Assume now to replicate system~\eqref{eq:sys} into $n$ independent copies, hence yielding a (tensorial) system
\begin{equation}
\label{eq:sysn}
 \frac{d\mathbf{x}^{(i)}}{dt}(t)=\mathbf{f}(\mathbf{x}^{(i)}(t))\quad\forall i=1,\dots, n\, ,
\end{equation}
where $\mathbf{x}^{(i)}(t)=(x^{(i)}_1(t),\dots,x^{(i)}_d(t))^\top$ denotes the state of the $i$-th copy of the generalised system. The whole system will thus be described by the $n\times d$ vector $\mathbf{x}=(\mathbf{x}^{(1)},\dots,\mathbf{x}^{(n)})^\top$. Finally we allow each system~\eqref{eq:sysn} to simultaneously interact with many others, and specifically belonging to the same hyperedge.

Let thus $E_\alpha$ be an hyperedge containing the $i$-th system. Then the growth rate associated to this latter will depend on all the systems $j \neq i$, belonging to the same hyperedge; moreover we assume such interaction to depend also on the hyperedge size, $\varphi(C_{\alpha\alpha})$, for a generic function $\varphi$. The system $i$ may belong to several hyperedges $E_\alpha$ and thus all these contributions should be taken 
into account to determine its growth rate. In formula
\begin{equation}
\label{eq:sysnhg}
 \frac{d\mathbf{x}^{(i)}}{dt}(t)=\frac{\sum_{\alpha} e_{i\alpha}\sum_j e_{j\alpha} \varphi(C_{\alpha\alpha}) \mathbf{F}(\mathbf{x}^{(i)}(t),\mathbf{x}^{(j)}(t))}{\sum_{\alpha} e_{i\alpha}\sum_j e_{j\alpha} \varphi(C_{\alpha\alpha})} \quad\forall i=1,\dots, n\, ,
\end{equation}
where we introduced the function $\mathbf{F}$ such that $\mathbf{F}(\mathbf{x}^{(i)},\mathbf{x}^{(i)})=\mathbf{f}(\mathbf{x}^{(i)})$ and the term at the denominator acts as a normalisation factor. We will show later on, that different functions $\mathbf{F}$ can be used to return the same function $\mathbf{f}$. 

Let us define the $m\times m$ diagonal matrix $\Phi$ such that $\Phi_{\alpha\alpha}=\varphi(C_{\alpha\alpha})$ and zero otherwise. Then we can rewrite Eq.~\eqref{eq:sysnhg} as follows
\begin{equation}
\label{eq:sysnhg2}
 \frac{d\mathbf{x}^{(i)}}{dt}(t)=\frac{1}{d_i}\sum_j D_{ij} \mathbf{F}(\mathbf{x}^{(i)}(t),\mathbf{x}^{(j)}(t)) \quad\forall i=1,\dots, n\, ,
\end{equation}
where we introduced the matrix $\mathbf{D}=\mathbf{e} \,\Phi \,\mathbf{e}^{\top}$ whose 
elements read
\begin{equation}
\label{eq:matrixD}
D_{ij} = \sum_{\alpha} e_{i\alpha}\Phi_{\alpha\alpha}e_{j\alpha}\quad \forall i\neq j\text{ and }D_{ii}=\varphi(1) \, .
\end{equation}
Let us observe that the different definition for the diagonal elements is 
due to the inclusion of the trivial hyperedges containing each single node and thus having size $1$. Finally let use define $d_i=\sum_j D_{ij}$.

\begin{remark}[Isolated systems]
 \label{rem:isol}
 In the case $n$ systems are isolated, i.e. all the hyperedges have size $1$, then $C_{\alpha\alpha}=1$ for all $\alpha$. Observing that a single $\alpha'$ (the one associated to the unique hyperedge containing $i$) does satisfy $e_{i\alpha'}=1$ (all the other ones being zero, $e_{i\beta}=0$ for all $\beta=\alpha'$), we can rewrite equation~\eqref{eq:sysnhg} by remarking that the sum over $j$ is restricted to $j=i$:
 \begin{equation*}
 \frac{d\mathbf{x}^{(i)}}{dt}(t)=\frac{\varphi(1) \mathbf{F}(\mathbf{x}^{(i)}(t),\mathbf{x}^{(i)}(t))}{\varphi(1)}= \mathbf{f}(\mathbf{x}^{(i)}(t))\quad\forall i=1,\dots, n\, ,
\end{equation*}
where use has been made of the relation $\mathbf{F}(\mathbf{x}^{(i)},\mathbf{x}^{(i)})= \mathbf{f}(\mathbf{x}^{(i)})$. Because our formalism contains the trivial case of isolated systems~\eqref{eq:sysn}, it results thus a natural extension of the latter.
\end{remark}

\begin{remark}[Pairwise interacting systems]
 \label{rem:pairwise}
 In case of systems interacting in pairs, i.e. when all hyperedges have size $C_{\alpha\alpha}=2$ for all $\alpha$ (but the ones associated to the trivial hyperedges containing each node), we can show that equation~\eqref{eq:sysnhg} converges back to the usual setting of a dynamical model anchored on a conventional network~\cite{CENCETTI2020109707}, once we assume $\varphi\equiv 1$, namely the same unitary weight is associated to each link.
 
First of all, let us observe that $D_{ii}=(\mathbf{e}\,\Phi \,\mathbf{e}^{\top})_{ii}= 
\varphi(1){A}_{ii}$ while  for $i\neq j$ we have $D_{ij}=(\mathbf{e}\,\Phi \,\mathbf{e}^{\top})_{ij}= \varphi(2) {A}_{ij}$, where we used the definition of the adjacency matrix that includes self-loops. Then Eq.~\eqref{eq:sysnhg2} can be rewritten as
 \begin{equation*}
 \frac{d\mathbf{x}^{(i)}}{dt}(t)=\frac{\sum_j A_{ij} \mathbf{F}(\mathbf{x}^{(i)}(t),\mathbf{x}^{(j)}(t))}{k_i}\quad\forall i=1,\dots, n\, ,
\end{equation*}
where use has been made of the definition $k_i=\sum_j A_{ij}$.
\end{remark}

\subsection{Dynamical behaviour}
\label{ssec:dynb}
Assume $\mathbf{s}(t)$ to be a solution of the initial system~\eqref{eq:sys}, then $\mathbf{x}^{(i)}(t)=\mathbf{s}(t)$, $i=1,\dots,n$, is trivially also a homogeneous solution of Eq.~\eqref{eq:sysn} but also of Eq.~\eqref{eq:sysnhg2}. Indeed, for all $i=1,\dots, n$ one has
\begin{eqnarray}
\label{eq:homsol}
 \frac{d\mathbf{x}^{(i)}}{dt}(t)&=&\frac{1}{d_i} {\sum_j D_{ij} \mathbf{F}(\mathbf{x}^{(i)}(t),\mathbf{x}^{(j)}(t))}\Big\rvert_{\mathbf{x}^{(i)}(t)=\mathbf{s}(t)}=\frac{1}{d_i}{\sum_j D_{ij} \mathbf{F}(\mathbf{s}(t),\mathbf{s}(t))}\notag\\
 &=&\frac{1}{d_i}{\sum_j D_{ij} \mathbf{f}(\mathbf{s}(t))}=\mathbf{f}(\mathbf{s}(t))\, ,
\end{eqnarray}
where we used the property $\mathbf{F}(\mathbf{s},\mathbf{s})=\mathbf{f}(\mathbf{s})$ and the definition of $d_i$. By definition of $\mathbf{s}$ 
the rightmost term equals $\dot{\mathbf{s}}$ which thus coincides also with the leftmost term.

Consider now a spatially dependent perturbation, i.e. a node depending one, about the homogeneous solution, $\mathbf{x}^{(i)}(t)=\mathbf{s}(t)+\mathbf{u}^{(i)}(t)$. Insert this ansatz into Eq.~\eqref{eq:sysnhg2} and determine the evolution of $\mathbf{u}^{(i)}(t)$ by assuming it to be small (i.e. using a first order expansion), $\forall i=1,\dots, n$:
\begin{eqnarray*}
 \frac{d\mathbf{u}^{(i)}}{dt}(t)+\frac{d\mathbf{s}}{dt}(t)&=&\frac{1}{d_i}{\sum_j D_{ij} \mathbf{F}(\mathbf{s}+\mathbf{u}^{(i)},\mathbf{s}+\mathbf{u}^{(j)})} \\&=&\mathbf{f}(\mathbf{s})+\frac{1}{d_i}{\sum_j D_{ij} \left(\sum_{\ell}\partial_{x_{\ell}^{(i)}}\mathbf{F}(\mathbf{s},\mathbf{s}){u}_{\ell}^{(i)}+\sum_{\ell}\partial_{{x}_{\ell}^{(j)}}\mathbf{F}(\mathbf{s},\mathbf{s}){u}_{\ell}^{(j)}\right)}\\
 &=&\mathbf{f}(\mathbf{s})+\sum_{\ell}\partial_{x_{\ell}^{(i)}}\mathbf{F}(\mathbf{s},\mathbf{s}){u}_{\ell}^{(i)}+\frac{1}{d_i}{\sum_j D_{ij} \sum_{\ell}\partial_{{x}_{\ell}^{(j)}}\mathbf{F}(\mathbf{s},\mathbf{s}){u}_{\ell}^{(j)}}\\
 &=&\mathbf{f}(\mathbf{s})+\mathbf{J}_1\mathbf{u}^{(i)}+\frac{1}{d_i}{\sum_j D_{ij} \mathbf{J}_2\mathbf{u}^{(j)}}\, ,
\end{eqnarray*}
where we defined the Jacobian matrices $\mathbf{J}_1=\partial_{\mathbf{x}_1} \mathbf{F}(\mathbf{s},\mathbf{s})$, i.e. the derivatives are computed with respect to the first group of variables, and $\mathbf{J}_2=\partial_{\mathbf{x}_2} \mathbf{F}(\mathbf{s},\mathbf{s})$, i.e. the derivatives are performed with respect to the second group of variables. In both cases the derivatives are evaluated at the reference solution $\mathbf{s}$.

By using the fact that $\dot{\mathbf{s}}=\mathbf{f}(\mathbf{s})$ and by 
slightly rewriting the previous equation, we obtain
\begin{eqnarray*}
 \frac{d\mathbf{u}^{(i)}}{dt}(t)&=&\mathbf{J}_1\mathbf{u}^{(i)}+\frac{1}{d_i}{\sum_j D_{ij} \mathbf{J}_2\mathbf{u}^{(j)}}=\mathbf{J}_1\mathbf{u}^{(i)}+\mathbf{J}_2\mathbf{u}^{(i)}+\sum_j \left(\frac{D_{ij}}{d_i}-\delta_{ij}\right) \mathbf{J}_2\mathbf{u}^{(j)}\\&=&\left(\mathbf{J}_1+\mathbf{J}_2\right)\mathbf{u}^{(i)}+\sum_j \mathcal{L}_{ij} \mathbf{J}_2\mathbf{u}^{(j)}
 \, ,
\end{eqnarray*}
where we defined the matrix operator 
\begin{equation}
\label{eq:consLapHG}
\mathcal{L}_{ij}=\frac{D_{ij}}{d_i}-\delta_{ij}\, .
\end{equation}

By introducing the $n\times d$ vector $\mathbf{u}=(\mathbf{u}^{(1)},\dots,\mathbf{u}^{(n)})^{\top}$ we can rewrite the latter equation in a compact form as:
\begin{equation}
\label{eq:finalEq}
 \frac{d\mathbf{u}}{dt}(t)=\left[\left(\mathbf{J}_1+\mathbf{J}_2\right)\otimes \mathbf{I}_n+\mathbf{J}_2 \otimes \mathcal{L} \right]\mathbf{u} \, ,
\end{equation}
where $\mathbf{I}_n$ is the $n\times n$ identity matrix and $\otimes$ is the Kronecker product of matrices.

One can prove that $\mathcal{L}$ is a novel (consensus) high-order Laplace matrix~\footnote{Let us introduce $\mathcal{L}^{\mathit{sym}}=\mathbf{d}^{-1/2}\mathbf{L}^H\mathbf{d}^{-1/2}$, where $\mathbf{d}$ is the diagonal matrix containing the $d_i$'s on the diagonal and $\mathbf{L}^H$ is the high-order (combinatorial) Laplace matrix defined in~\cite{CarlettiJPC_2020}. Then $\mathcal{L}^{\mathit{sym}}=D_{ij}/\sqrt{d_i d_j} -\delta_{ij}$ from which it immediately follows that $\mathcal{L}^{\mathit{sym}}$ 
is symmetric and nonpositive definite; indeed take any $\mathbf{x}\in\mathbb{R}^N\setminus \{0\}$, $N$ standing for the dimension of the matrices, 
then $(\mathbf{x},\mathcal{L}^{\mathit{sym}} \mathbf{x})=(\mathbf{d}^{-1/2} \mathbf{x},\mathbf{L}^H\mathbf{d}^{-1/2} \mathbf{x})\leq 0$ where the last inequality follows from the fact that $\mathbf{L}^H$ is nonpositive definite. Finally let us observe that $\mathcal{L}=\mathbf{d}^{-1}\mathbf{L}^H =\mathbf{d}^{-1/2}\mathcal{L}^{\mathit{sym}}\mathbf{d}^{1/2}$, hence, $\mathcal{L}$ is similar to $\mathcal{L}^{\mathit{sym}}$ and, thus they display the same non-positive spectrum. Moreover this implies also that $-2\leq \Lambda^{(\alpha)}\leq 0$.} , i.e. it is nonpositive definite, the largest eigenvalue is $\Lambda^{(1)}=0$ and its is associated to the uniform eigenvector $\phi^{(1)}\sim (1,\dots,1)^{\top}$.

Recalling the relation $\mathbf{f}(\mathbf{x})=\mathbf{F}(\mathbf{x},\mathbf{x})$ one can prove that:
\begin{equation*}
\partial_{\mathbf{x}}\mathbf{f}:= \mathbf{J}=\mathbf{J}_1+\mathbf{J}_2\, ,
\end{equation*}
and thus rewrite Eq.~\eqref{eq:finalEq} as
\begin{equation}
\label{eq:finalEq2}
 \frac{d\mathbf{u}}{dt}(t)=\left[\mathbf{J}\otimes \mathbf{I}_n+\mathbf{J}_2 \otimes \mathcal{L} \right]\mathbf{u} \, .
\end{equation}

This is a linear system involving matrices with size $nd\times nd$. To progress with the analytical understanding, we employ the eigenbase of $\mathcal{L}$, to project the former equation onto each eigendirection
\begin{equation}
\label{eq:finalEqalpha}
 \frac{d\mathbf{u}^{(\alpha)}}{dt}(t)=\left[\mathbf{J}(\mathbf{s}(t))+\mathbf{J}_2 (\mathbf{s}(t))\Lambda^{(\alpha)}\right]\mathbf{u}^{(\alpha)} 
\, ,
\end{equation}
where $\Lambda^{(\alpha)}$ is the eigenvalue relative to the eigenvector $\phi^{(\alpha)}$. The above equation enables us to infer the stability of the homogeneous solution, $\mathbf{s}(t)$, by studying the Master Stability Function, namely the real part of the largest Lyapunov exponent of Eq.~\eqref{eq:finalEqalpha}. To illustrate the potentiality of the theory we shall turn to considering a specific application that we will introduce in the following.

\subsection{Results}
\label{ssec:reshhcoup}
In the above analysis we have obtained a one-parameter family (indexed by 
the eigenvalues $\Lambda^{(\alpha)}$) of linear but (in general) time dependent systems~\eqref{eq:finalEqalpha}. For the sake of simplicity we will hypothesise the homogenous solution to be stationary and stable, $\mathbf{s}(t)=\mathbf{s}_0$. In this way we will hence assume each isolated system to converge to the same stationary point. This simplifies the study of Eq.~\eqref{eq:finalEqalpha}, by allowing us to deal with a constant linear system. Let us observe that one could in principle study the more general setting of a time dependent solution, by using the Floquet theory 
in case of a periodic orbit or the full Master Stability Function in the case of irregular oscillators.

As a concrete application we will consider a Volterra model~\cite{mckane2005predator} which describes the interaction of prey and predators in an ecological setting :
\begin{equation}
\label{eq:Voltnospace}
\begin{cases}
 \dot{x}=- d x+c_1 xy \\
 \dot{y}=ry - sy^2 - c_2xy\, ,
 \end{cases}
\end{equation}
here $x$ denotes the concentration of predators, while $y$ stands for the 
prey and $\dot{ }$ the time derivative. All the parameters are assumed to 
be positive; in the following we will make use of the choice $c_1 = 2$, 
$c_2 = 13$, $r = 1$, $s = 1$ and $d = 1/2$, but of course our results hold true in general. The Volterra model~\eqref{eq:Voltnospace} admits a nontrivial fixed-point, $x^* = \frac{c_1r-sd}{c_1c_2}$, $y^*=\frac{d}{c_1}$, which is positive and stable, provided $c_1r-sd>0$. In the case under scrutiny, we have $x^*=3/52\sim 0.0577$ and $y^*=1/4$.

Following the above presented scheme, let us now considering $n$ replicas 
of the model~\eqref{eq:Voltnospace}, each associated to a different ecological niche and indexed by the node index $i$. Assume also
that species can sense the remote interaction with other communities populating  neighbouring nodes. For instance, the competition of prey for food and resources can be easily extended so as to account for a larger habitat which embraces adjacent patches. At the same time, predators can benefit from a coordinated action to hunt in team. For a sake of definitiveness we will study in the following the high-order coupling (let us stress once again that several ``microscopic'' high-order models can give rise to the same network-aggregate model) defined by:
\begin{equation}
\label{eq:VoltnospaceHG}
\begin{cases}
 \dot{x}_i=- d x_i+a c_1 y_i \frac{1}{d_i}\sum_j D_{ij}x_j +(1-a) c_1 x_i  \frac{1}{d_i}\sum_j D_{ij} y_j \\
 \dot{y}_i=ry_i - sy_i \frac{1}{d_i}\sum_j D_{ij}y_j - c_2y_i \frac{1}{d_i}\sum_j D_{ij}x_j \, ,
 \end{cases}
\end{equation}
where the matrix $D_{ij}$ encodes for the high-order interaction among nodes $i$ and $j$, taking into account the number and size of the hyperedges containing both nodes
 (see~\eqref{eq:matrixD}). The parameters $a \in[0,1]$ describes the relative strength with which the predators in node $i$ increase because of the ``in-node'' predation or because of the interaction among predators in the hyperedges. The case $a=1$ corresponds to a purely in-node process while if $a=0$ a coordinated action to hunt in team is assumed to rule the dynamics. Preys feel the competition for the resources with preys living in nodes belonging to the same hyperedge (second term on the right hand side of the second equation of~\eqref{eq:VoltnospaceHG}) as well from predators in the same hyperedge (rightmost terms in the same equation). Birth and death of both species are local, i.e. due to resources available 
in-node.

By using the new Laplace matrix~\eqref{eq:consLapHG} we can rewrite the previous model~\eqref{eq:VoltnospaceHG} as:
\begin{equation}
\label{eq:VoltnospaceHG2}
\begin{cases}
 \dot{x}_i=- d x_i+c_1 y_i x_i +a c_1 y_i \sum_j \mathcal{L}_{ij}x_j +(1-a) c_1 x_i  \sum_j \mathcal{L}_{ij} y_j \\
 \dot{y}_i=ry_i - sy^2_i - c_2y_i x_i-sy_i \sum_j \mathcal{L}_{ij}y_j - 
c_2y_i \sum_j \mathcal{L}_{ij}x_j \, ,
 \end{cases}
\end{equation}
where one can easily recognise the in-node Volterra model~\eqref{eq:Voltnospace} and the corrections stemming from high-order contributions.

\begin{figure}[t]
\sidecaption[t]
\includegraphics[scale=.30]{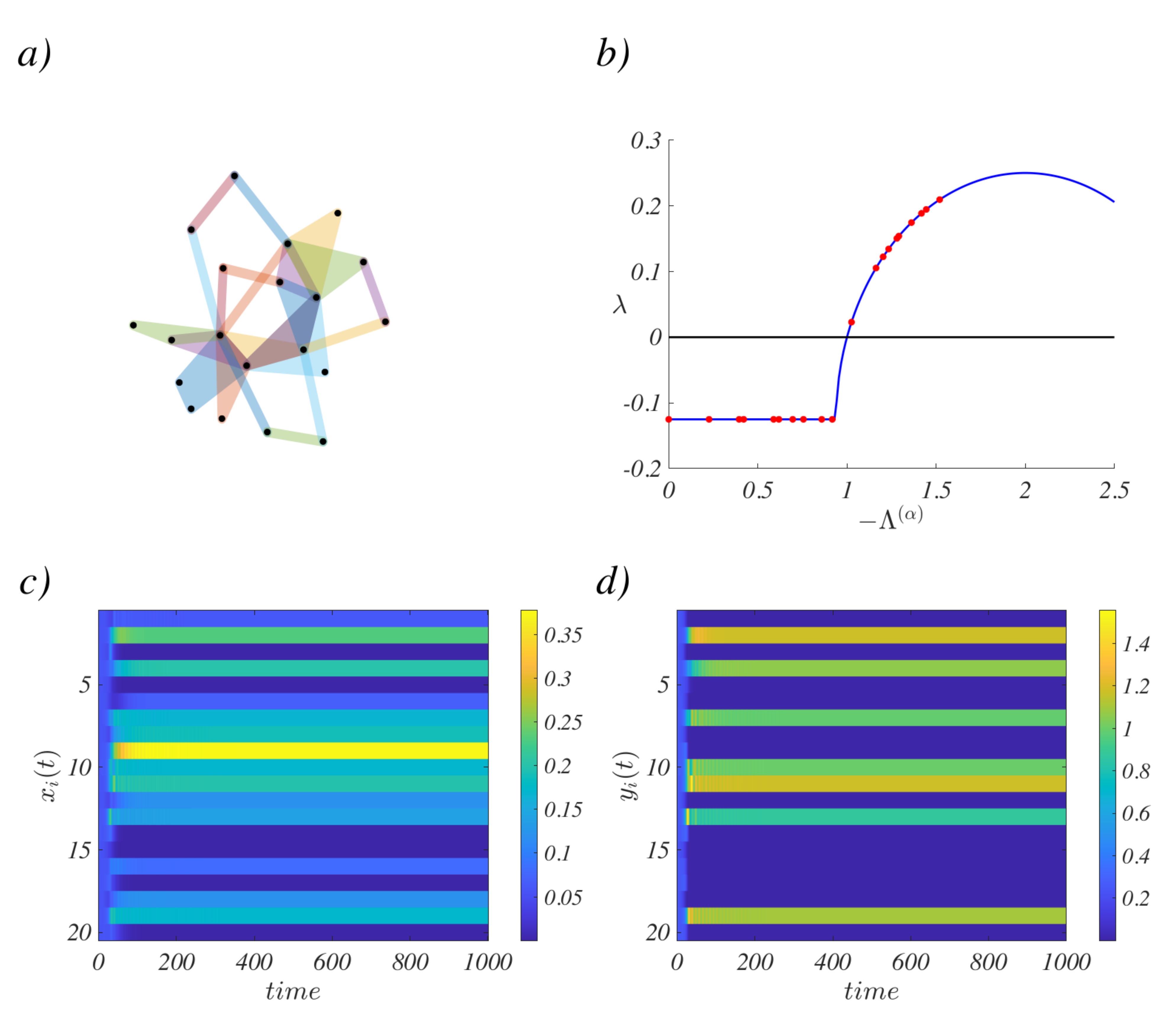}
%
%
\caption{\textbf{Patterns in the Volterra model with high-order interactions (I)}. In panel a) we represent the hypergraph used to model the high-order interactions among species living in different niches. The hypergraph is composed of $n=20$ nodes and it has been generated using a random 
attachment process and it is composed by $20$ trivial hyperedges of size $1$, $11$ hyperedges of size $2$, $10$ hyperedges of size $3$ and $1$ hyperedge of size $4$. In panel b) we report the dispersion relation for the 
Volterra model~\eqref{eq:VoltnospaceHG}, the red symbols refer to $\lambda\left(\Lambda^{(\alpha)}\right)$, $\alpha\in\{1,\dots,n\}$, while the blue line denotes the dispersion relation for the Volterra model reformulated on a continuous support. In panel c) we show the time evolution of the 
predator density in each node as a function of time, $x_i(t)$; let us observe that in (almost) each node the density of predators is much larger than the corresponding homogenous equilibrium $x^*\sim 0.0577$ (blue). Panel d) report the time evolution of the prey density in each node as a function of time, $y_i(t)$; let us observe that in (almost) each node the density of preys is much lower than the corresponding homogenous equilibrium $y^*= 1$ (green). The model parameters have been set to $c_1 = 2$, $c_2 = 13$, $r = 1$, $s = 1$, $d = 1/2$ and $a=1/2$. We fix $\varphi(c)=c^\sigma$ with $\sigma=1.5$.}
\label{fig:PattIntHG}       
\end{figure}

As previously shown, in the general setting (see~\eqref{eq:homsol}) the homogenous solution $(x^*,y^*)$ is also a solution of the coupled system~\eqref{eq:VoltnospaceHG}, that is $x_i=x^*$ and $y_i=y^*$ solves the latter. In the following we will prove that such solution can be destabilised due to the high-order coupling so driving the system towards a new heterogenous, spatially dependent, solution. To prove this claim, we will linearise system~\eqref{eq:Voltnospace} about the homogeneous equilibrium by setting $u_i=x_i-x^*$ and $v_i=y_i-y^*$ and then make use of the eigenbase of the Laplace matrix $\mathcal{L}$, $(\Lambda^{(\alpha)},\phi^{(\alpha)})$, to project the linear system onto each eigenmode, that is $u_i=\sum_\alpha u^{\alpha}\phi_i^{(\alpha)}$ and $v_i=\sum_\alpha v^{\alpha}\phi_i^{(\alpha)}$:
\begin{eqnarray}
\label{eq:VoltnospaceHG2lin}
\frac{d}{dt}\binom{u^\alpha}{v^\alpha}&=&\left[\left(
\begin{matrix}
 0 & c_1 x^*\\
 -c_2y^* & -sy^*
\end{matrix}
\right)+\Lambda^{(\alpha)}\left(
\begin{matrix}
 ac_1y^* & (1-a)c_1 x^*\\
 -c_2y^* & -sy^*
\end{matrix}
\right)\right]\binom{u^\alpha}{v^\alpha}\notag\\
&=& \left(\mathbf{J}+\Lambda^{(\alpha)} \mathbf{J}_2\right)\binom{u^\alpha}{v^\alpha} =:\mathbf{J}^{(\alpha)} \binom{u^\alpha}{v^\alpha}\, .
\end{eqnarray}
The homogenous solution will prove unstable if (at least) one eigenmode $\bar{\alpha}$ exists for which the largest real part of the eigenvalues of $\mathbf{J}^{(\bar{\alpha})}$ is positive. The real part of the largest 
eigenvalue $\lambda$ as function of $\Lambda^{(\alpha)}$ is called the {\em dispersion relation}. One can easily realise that $\lambda$ is the solution with the largest real part of the second order equation
\begin{equation*}
 \lambda^2-\mathrm{tr}\mathbf{J}^{(\alpha)}\lambda+\det \mathbf{J}^{(\alpha)}=0\, .
\end{equation*}
Hence the required condition for the instability is 
\begin{equation}
\label{eq:instab}
\mathrm{tr}\mathbf{J}^{(\alpha)}>0 \text{ or }\mathrm{tr}\mathbf{J}^{(\alpha)}<0 \text{ and }\det\mathbf{J}^{(\alpha)}<0\, .
\end{equation}
A straightforward computation returns
\begin{equation*}
 \mathrm{tr}\mathbf{J}^{(\alpha)}=-sy^*+\Lambda^{(\alpha)}\left(-s+ac_1\right) \text{ and }\det\mathbf{J}^{(\alpha)}=c_1y^*\left(1+\Lambda^{(\alpha)}\right)\left[\Lambda^{(\alpha)}\left(c_2x^*(1-a)-asy^*\right)+c_2x^*\right]\, .
\end{equation*}
Let us recall that the homogenous equilibrium is stable for the decoupled 
system corresponding to setting $\Lambda^{(1)}=0$. Indeed $\mathrm{tr}\mathbf{J}^{(1)}=-sy^*<0$ and $\det\mathbf{J}^{(1)}=c_1c_2x^*y^*>0$. We have thus to determine the existence of (at least one) $\bar{\alpha}\geq 2$ for which the conditions for instability~\eqref{eq:instab}, allowing 
us to prove the positivity of $\lambda\left(\Lambda^{(\bar{\alpha})}\right)$. In Fig.~\ref{fig:PattIntHG} we report a case where the high-order coupling is able to destabilise the homogenous solution (panel b), thus returning a patchy solution (panels c and d) for the involved species. Finally let us observe that interestingly some niches ($6$ over $20$) become empty, that is deprived of any species.

\begin{figure}[t]
\sidecaption[t]
\includegraphics[scale=.30]{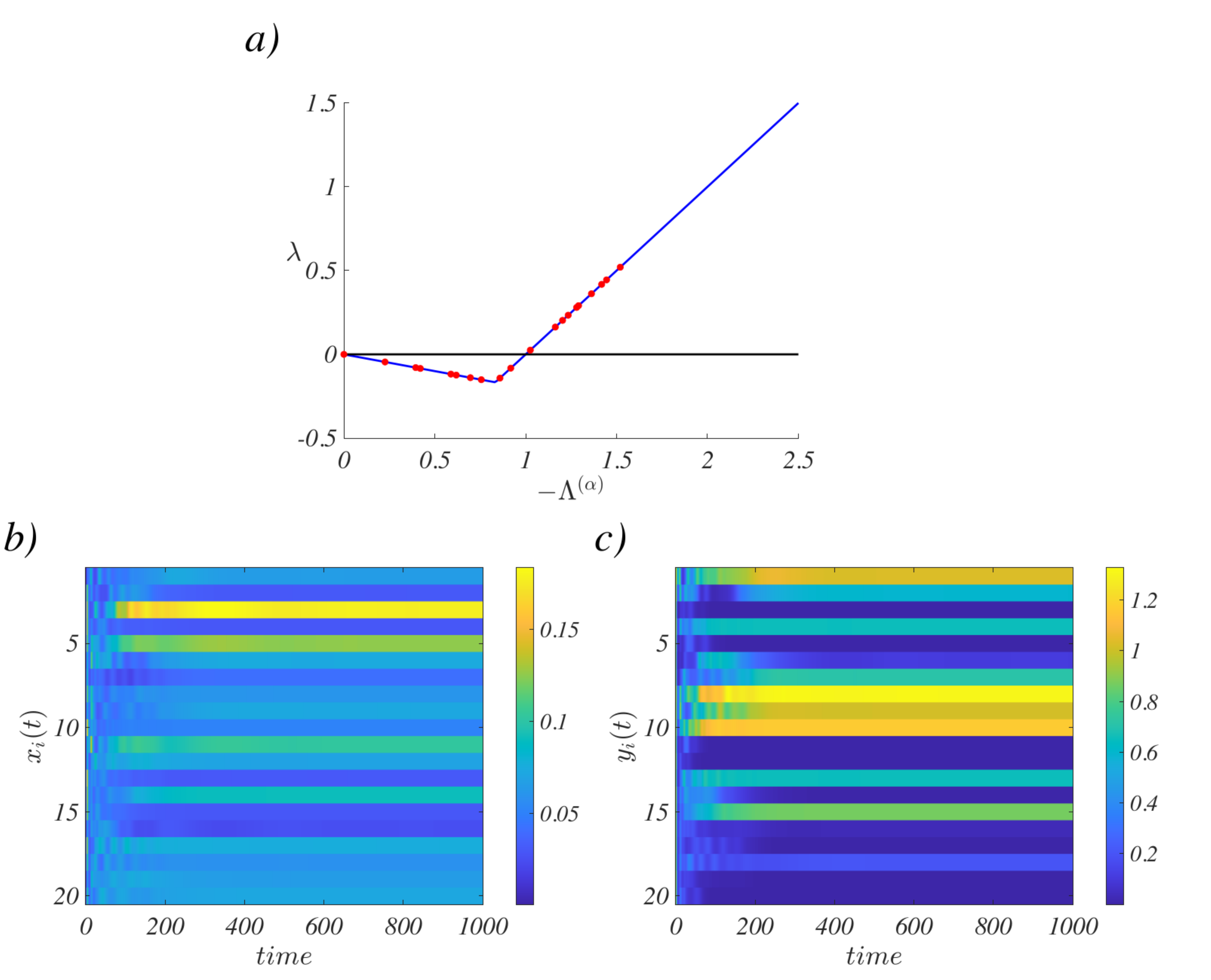}
%
%
\caption{\textbf{Patterns in the Volterra model with high-order interactions (II)}. Using the same hypergraph shown in Fig.~\ref{fig:PattIntHG} we 
study the emergence of patterns close to the homogeneous equilibrium $\tilde{x}=0$ and $\tilde{y}=r/s=1$. We report in panel a) the dispersion relation for the Volterra model~\eqref{eq:VoltnospaceHG}, the red symbols refer to $\lambda\left(\Lambda^{(\alpha)}\right)$, $\alpha\in\{1,\dots,n\}$, while the blue line denotes the dispersion relation for the Volterra model computed on a continuous support. In panel b) we show the time evolution of the predator density in each node as a function of time, $x_i(t)$; let us observe that in each node the density of predators is positive in striking contrast with it happens for the uncoupled system. Panel c) reports the time evolution of the prey density in each node as a function of time, $y_i(t)$; let us observe that in each node the density of preys is much lower than the homogenous equilibrium $y^*= 1$ (green) and in $8$ niches the preys have gone through extinction. The model parameters have been set to $c_1 = 2$, $c_2 = 13$, $r = 1$, $s = 1$, $d = 
1/2$ and $a=1/2$. We fix $\varphi(c)=c^\sigma$ with $\sigma=1.5$.}
\label{fig:PattIntHG0}       
\end{figure}

Another even more interesting case is reported in Fig.~\ref{fig:PattIntHG0}. In this case the uncoupled homogeneous equilibrium yields $\tilde{x}=0$ and $\tilde{y}=r/s$. When extending the study to account for multi body interactions, predators do survive in each niche while the preys go through extinction in a few location ($6$ nodes over $20$). Generally the density of preys is lower than the equilibrium value found in the isolated case.

\section{Hypergraph and high-order diffusive-like coupling}
\label{sec:lap}
In the previous section we have introduced and studied the problem of the 
emergence of a spatially heterogenous solution in a system of several identical dynamical units coupled together via the (hyper) adjacency matrix of the hypergraph. In particular the microscopic units defining the system are constrained to stay anchored to the node where they interact with those sharing the same location and those belonging to nodes of the incident hyperedges. In this section we will present a modified framework based 
on the assumption that the basic units can travel across the hypergraph jumping from node to node via the available hyperedges.

Starting from the definition of hyper adjacency matrix, Eq.~\eqref{eq:hyperadj}, the notion of (combinatorial) Laplace matrix for networks can be straightforwardly generalised to the case of hypergraphs~\cite{JM2019,MKJ2020}, by defining  $k_i\delta_{ij}-A_{ij}$, where $k_i=\sum_j A_{ij}$. 
Let us however observe that the latter does not account in full for the higher-order structures encoded in the hypergraph. Notably, the sizes of the incident hyperedges are neglected. 

To overcome this limitation, authors of~\cite{CarlettiEtAl2020}  studied a random walk process defined on a generic hypergraph using a new (random 
walk) Laplace matrix. It is worth mentioning that the transition rates of 
the associated process, linearly correlates with the size of the involved 
hyperedges. Stated differently, exchanges are favoured among nodes belonging to the same hyperedge (weighted according to its associated size). Note that a similar construction has been proposed in~\cite{Evans_2010} to extract a $n$-clique graph from a network. The main difference in the present case is that hyperedges can have an heterogeneous size distribution and thus provide a more flexible framework for tackling a wide range of problems.

For the sake of completeness, let us briefly recall the construction of the random walk process on a hypergraph and invite the interested reader to consult~\cite{CarlettiEtAl2020} for further details. The agents are located on the nodes and hop between them. In a general setting, the walkers 
may weight hyperedges depending on their size, introducing a bias in their moves that we shall encode into a function $\varphi$ of the hyperedge size. This yields the weighted adjacency matrix $\mathbf{D}=\mathbf{e}\,\Phi\,\mathbf{e}^{\top}$, already defined in Eq.~\eqref{eq:matrixD} and hereby recalled:
\begin{equation*}
D_{ij} = \sum_{\alpha} e_{i\alpha}\Phi_{\alpha\alpha}e_{j\alpha}\quad \forall i\neq j\text{ and }D_{ii}=\varphi(1) \, ,
\end{equation*}
where $\Phi$ is the diagonal matrix whose elements read $\varphi(C_{\alpha\alpha})$.
The transition probabilities of the examined process are then obtained by 
normalising the columns of the weighted adjacency matrix $T_{ij}=\frac{D_{ij}}{d_i}$ for all $i$, where again $d_i=\sum_{j} D_{ij}$.

Let us briefly observe that assuming $\varphi(c)=c^\sigma$ allows to cover several existing models of random walks on hypergraphs.
For $\sigma=1$, we  get the random walk defined in~\cite{CarlettiEtAl2020}, while for $\sigma=-1$ we obtain the one introduced by Zhou~\cite{zhou2007learning}. Finally, the case $\sigma=0$ returns a random walk on 
the so called {\em clique reduced multigraph}. The latter is a multigraph 
where each pair of nodes is connected by a number of edges equal to the number of hyperedges containing that pair in the hypergraph.

From the above introduced transition probabilities one can define the {{\em random walk}} Laplacian generalising that of standard networks, $L_{ij}=  \delta_{ij}-T_{ij}$, and eventually derive the (combinatorial) Laplace matrix, 
\begin{equation}
\label{eq:Laphg}
\mathbf{L}^H=\mathbf{D}-\mathbf{d}\, ,
\end{equation}
this latter will be employed in the following to model diffusion on higher-order structures. In the above equation, matrix $\mathbf{d}$ displays, on the diagonal, the values $d_i=\sum_{j} D_{ij}$ and zeros otherwise. It is clear from its very definition that $\mathbf{D}$ takes into account 
both the number and the size of the hyperedges incident with the nodes. It can also be noted that $\mathbf{D}$ can be considered as a {\em weighted adjacency matrix} whose weights have been self-consistently defined so as to account {for} 
the higher-order structures encoded in the hypergraph.

Consider again the $d$-dimensional system Eq.~\eqref{eq:sys} described by 
local, i.e. aspatial, equations:
\begin{equation}
\label{eq:dotxF}
\frac{d\mathbf{x}}{dt}=\mathbf{f}(\mathbf{x})\quad \mathbf{x}\in\mathbb{R}^d\, ,
\end{equation}
and assume further $n$ identical copies of the above system coupled through a hypergraph. In this way each copy of the system attached to a node of a hypergraph belonging to one (or more) hyperedge. Units sharing the same hyperedge are tightly coupled, due to existing many body interactions. 
In formulas:
\begin{equation*}
\frac{d\mathbf{x}_i}{dt}=\mathbf{f}(\mathbf{x}_i) +\varepsilon \sum_{\alpha:i\in E_\alpha}\sum_{j\in E_\alpha}\varphi(C_{\alpha\,\alpha})\left(\mathbf{G}(\mathbf{x}_j)-\mathbf{G}(\mathbf{x}_i)\right)\, ,
\end{equation*}
where $\mathbf{x}_i$ denotes the state of the $i$-th unit, i.e. anchored to the $i$-th node, $\varepsilon$ the strength of the coupling, $\varphi$ 
is the function encoding the bias due to the hyperedge size and $\mathbf{G}$ a generic nonlinear coupling function. From the definition of $e_{i\alpha}$ one can rewrite the previous formula as
\begin{eqnarray}
\label{eq:maineq}
\frac{d\mathbf{x}_i}{dt}&=&\mathbf{f}(\mathbf{x}_i) +\varepsilon \sum_{\alpha,j} e_{i\alpha}e_{j\alpha}\varphi(C_{\alpha\,\alpha})\left(\mathbf{G}(\mathbf{x}_j)-\mathbf{G}(\mathbf{x}_i)\right)\notag\\
&=&\mathbf{f}(\mathbf{x}_i) +\varepsilon\sum_{j} D_{ij}\left(\mathbf{G}(\mathbf{x}_j)-\mathbf{G}(\mathbf{x}_i)\right)=\mathbf{f}(\mathbf{x}_i) 
+\varepsilon\sum_{j} \left(D_{ij}-d_{i}\delta_{ij}\right)\mathbf{G}(\mathbf{x}_j)\notag\\
&=&\mathbf{f}(\mathbf{x}_i) +\varepsilon\sum_{j} L^H_{ij}\mathbf{G}(\mathbf{x}_j)\, ,
\end{eqnarray}
where we have used the above definitions for $d_i$ and $L^H_{ij}$. Let us 
stress once again that the whole high-order structure is encoded in a $n\times n$ matrix. Hence there is no need for tensors and this simplifies the resulting analysis.

By exploiting the fact that $\sum_j L^H_{ij}=0$ for all $i=1,\dots, n$, it is immediate to conclude that the aspatial reference solution $\mathbf{s}(t)$, i.e. the time dependent function solving Eq.~\eqref{eq:dotxF}, is also a solution of Eq.~\eqref{eq:maineq}. A natural question hence arises: what can we say of the stability of the homogeneous solution for the system in its diffusive-like coupled variant?

To answer to this question one introduces again the deviations from the reference orbit, i.e. $\mathbf{u}_i=\mathbf{x}_i-\mathbf{s}$. Assuming this latter to be small, one can derive a self-consistent set of linear differential equations for tracking the evolution of the perturbation in time. To this end, we make use of the  expression in the above Eq.~\eqref{eq:maineq} and  perform a Taylor expansion to the linear order of approximation, to eventually get:
\begin{equation}
\label{eq:GLHGlin}
\frac{d\mathbf{u}_i}{dt}=\mathbf{J}(\mathbf{s}(t))\mathbf{u}_i +\varepsilon \sum_{j} {L}^H_{ij} \mathbf{J}_\mathbf{G}(\mathbf{s}(t))\mathbf{u}_j\, , 
\end{equation}
where $\mathbf{J}(\mathbf{s}(t))$ (resp. $ \mathbf{J}_\mathbf{G}(\mathbf{s}(t))$) denotes the Jacobian matrix of the function $\mathbf{f}$ (resp. $\mathbf{G}$) evaluated on the trajectory $\mathbf{s}(t)$.

We can improve on our analytical understanding of the problem by employing again the eigenbase of the Laplace matrix $\mathbf{L}^H$. Being the latter symmetric there exists a basis of orthonormal eigenvectors, $\phi_H^{(\alpha)}$, associated to the eigenvalues $\Lambda_H^{(\alpha)}$. We can then project $\mathbf{u}_i$ on this basis and obtain, for all $\alpha$:
\begin{equation}
\label{eq:GLHGlinalpha}
\frac{d\mathbf{y}_\alpha}{dt}=\left[\mathbf{J}(\mathbf{s}(t))+\varepsilon {\Lambda^{(\alpha)}_H} \mathbf{J}_\mathbf{G}(\mathbf{s}(t))\right]\mathbf{y}_\alpha\, , 
\end{equation}
where $\mathbf{y}_\alpha$ is the projection of $\mathbf{u}_i$ on the $\alpha$-th eigendirection. 

The (in)stability of the homogenous solution $\mathbf{s}(t)$ can be checked by looking at the eigenvalue of the linear system~\eqref{eq:GLHGlinalpha}, and more specifically the eigenvalue with the largest real part. In a general framework, where i.e. $\mathbf{s}(t)$ depends on time, we are dealing with a time dependent eigenvalue problem that can be tackled by using the Master Stability Function~\cite{Pecora,HCLP}. For  simplicity we will hereby solely consider the case of a stationary reference orbit, i.e. $\mathbf{s}(t)=\mathbf{s}_0$. In this way Eq.~\eqref{eq:GLHGlinalpha} can be directly solved by using spectral methods. We invite the interested reader to refer to~\cite{CarlettiJPC_2020} where the general case of 
a periodic or even a chaotic $\mathbf{s}(t)$ has been analysed.

\subsection{Turing patterns on hypergraphs}
\label{ssec:TP}

The problem introduced in the previous section opens up the perspective to address the notion of a Turing instability on hypergraphs. Indeed, according to the Turing instability mechanism, a stable homogeneous equilibrium becomes unstable upon injection of a heterogeneous, i.e. spatially dependent, {perturbation} once diffusion and reaction terms are simultaneously at play. The Turing phenomenon is exemplified with reference to $2$ dimensional systems. In the following we will consequently assume $d=2$ and rewrite $\mathbf{x}_i=(u_i,v_i)$ as well as $\mathbf{f}(\mathbf{x}_i)=\left(f(u_i,v_i),g(u_i,v_i)\right)$, where the index $i=1,\dots, n$ 
refers to the specific node to which the dynamical variables are bound. Hence Eq.~\eqref{eq:maineq} becomes
\begin{equation}
\label{eq:glodyn}
\begin{cases}
\dot{u_i} &=f(u_i,v_i)+D_u \sum_j L^H_{ij}u_j\\
\dot{v_i} &=g(u_i,v_i) +D_v \sum_j L^H_{ij}v_j
\end{cases}\, ,
\end{equation}
where $D_u$ and $D_v$ replace the diffusion coefficients of species $u$ and $v$ in the case of network and can thus be called generalised diffusion coefficients. At first sight, the above model seems to solely account for binary interactions. However, higher-order interactions are also present, as encoded in the matrix $\mathbf{L}^H$. Finally, let us observe that 
if the hypergraph is a network, then $\mathbf{L}^H$ reduces to the standard Laplace matrix and thus Eqs. (\ref{eq:glodyn}) converges to the usual reaction-diffusion system defined on a network. 

The {condition for the emergence of a} Turing instability can be assessed 
by performing a linear stability analysis about the homogeneous equilibrium~\cite{NM2010,Asllani2013,Asllani2014NC,Asllani2014}, as previously shown. Assuming $\mathbf{G}$ to be the identity function and the reference orbit to coincide with a stable stationary equilibrium $\mathbf{s}_0=(u_0,v_0)$, Eq.~\eqref{eq:GLHGlin} simplifies into:
 \begin{equation*}
\begin{cases}
\dot{\delta u_i} &=\partial_u f(u_0,v_0)\delta u_i +\partial_v f(u_0,v_0)\delta v_i +D_u \sum_j L^H_{ij}\delta u_j\\
\dot{\delta v_i} &=\partial_u g(u_0,v_0)\delta u_i +\partial_v g(u_0,v_0)\delta v_i +D_v \sum_j L^H_{ij}\delta v_j\, ,
\end{cases}
\end{equation*}
where $\delta u_i=u_i-u_0$ and $\delta v_i=v_i-v_0$. By exploiting again the eigenbasis of the Laplace matrix we can write $\delta u_i (t)= \sum_{\alpha} \hat{u}^\alpha(t) \phi^\alpha_i$ and $\delta v_i (t)= \sum_{\alpha} \hat{v}^\alpha(t) \phi^\alpha_i$. Finally the ansatz, $\hat{u}^\alpha(t) \sim e^{\lambda_\alpha t}$ and $\hat{v}^\alpha(t) \sim e^{\lambda_\alpha t}$, allows us to compute the dispersion relation, i.e. the linear growth rate  $\lambda_\alpha =\lambda(\Lambda_H^\alpha)$ of the eigenmode $\alpha$, as a function of the Laplacian eigenvalue $\Lambda_H^\alpha$.

{As it can be straightforwardly proved,} the linear growth rate is the largest real part of the roots of the second order equation
\begin{equation}
\label{eq:reldisp}
\lambda_\alpha^2 -\lambda_\alpha\left[ \mathrm{tr} \mathbf{J}_0+\Lambda_H^\alpha(D_u+D_v)\right]+\det \mathbf{J}_0+\Lambda_H^\alpha(D_u\partial_v g+D_v\partial_u f)+D_uD_v (\Lambda_H^\alpha)^2=0\, ,
\end{equation}
where $\mathbf{J}_0=\left(
\begin{smallmatrix}
 \partial_u f & \partial_v f\\
  \partial_u g & \partial_v g
\end{smallmatrix}
\right)$ is the Jacobian matrix of the reaction part evaluated at the equilibrium $(u_i,v_i)=(u_0,v_0)$. In Eq.~\eqref{eq:reldisp}, $\mathrm{tr}(\cdot)$ and $\det(\cdot)$ stand respectively for the trace and the determinant. The existence of at least one eigenvalue $\Lambda_H^{\tilde{\alpha}}$ for which the dispersion relation takes positive values, implies that the system goes unstable via a typical path first identified by Alan Turing in his seminal work. At variance, if the dispersion relation is negative the system cannot undergo a Turing instability: any tiny perturbation fades away and the system settles back to the homogeneous equilibrium.

\begin{figure}[t]
\sidecaption[t]
\includegraphics[scale=.30]{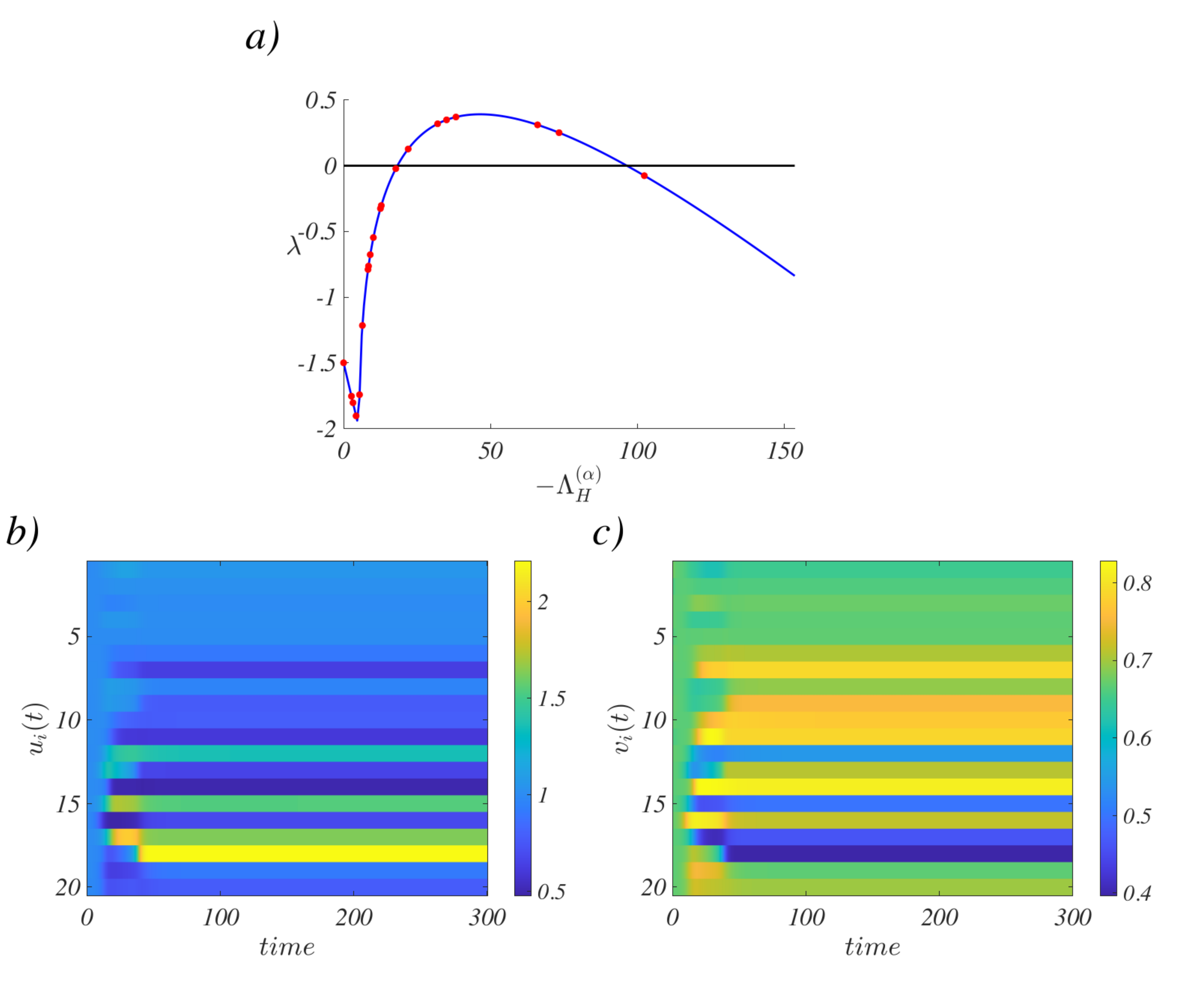}
%
%
\caption{\textbf{Turing patterns in the Brusselator model with high-order 
diffusive-like couplling}. Using the same hypergraph shown in Fig.~\ref{fig:PattIntHG} we study the   Turing patterns emerging from the homogeneous equilibrium $({u}_0,{v}_0)$. We report in panel a) the dispersion relation for the Brusselator model defined by the reaction terms $f(u,v)=1-(b+1)u+c u^2v$ and $g(u,v)=bu-cu^2v$; the red symbols refer to $\lambda\left(\Lambda_H^{(\alpha)}\right)$, $\alpha\in\{1,\dots,n\}$, while the blue line denotes the dispersion relation for the Brusselator model defined 
on a continuous support. In panel b) we show the time evolution of the $u$ variable in each node as a function of time, $u_i(t)$. Panel c) reports 
the time evolution of the $v$ variable in each node as a function of time, $v_i(t)$. The model parameters have been set to $b = 4$, $c= 6$, $D_u = 0.02$ and $D_v=0.17$. Hence $u_0=1$ and $v_0=b/c=2/3$. We fix $\varphi(c)=c^\sigma$ with $\sigma=1.5$.}
\label{fig:TPattIntHG}       
\end{figure}

To proceed further with a concrete example we selected the Brusselator reaction system~{\cite{PrigogineNicolis1967,PrigogineLefever1968}}. This is 
a nonlinear model defined by $f(u,v)=1-(b+1)u+c u^2v$ and $g(u,v)=bu-cu^2v$, where $b$ and $c$ act as tunable parameters. In Fig.~\ref{fig:TPattIntHG}  we report the results for a choice of the model parameters giving rise to Turing patterns ($b = 4$, $c = 6$, $D_u = 0.02$ and $D_v 
= 0.17$) and the same hypergraph previously used in Figs.~\ref{fig:PattIntHG} and~\ref{fig:PattIntHG0}. The dispersion relation (panel a) is clearly positive for a selection of $\Lambda_H^{(\alpha)}$ (red points). The 
homogeneous solution becomes hence unstable and the ensuing patterns are displayed in panels b) and c).

\section{Conclusions}
\label{sec:conc}
Complex systems are composed of a large number of simple units, mutually interacting via  nonlinear exchanges. Many-body interactions sit hence at the root of a large plethora of spontaneously emerging phenomena, as exhibited by complex systems. The former are often reduced to a vast collection of pairwise interactions, involving agents interacting in pairs. This enables one to model the inspected problem as a dynamical system flowing on a conventional binary network, a powerful approximation that allows for progresses to be made. In many cases of interest, this reductionist choice constitutes  a rough first order approximation to the examined dynamics and more precise models are to be invoked which encompass for the high-order interactions being at play.

In this work, we presented a general framework which allows one to account for multi-body interacting systems coupled via a hypergraph. This materialises in a natural extension of the conventional network paradigm. More specifically, we considered the problem of the emergence of heterogeneous stable solutions in interconnected systems, under the assumption that, once isolated, all  units converge to the 
same, and thus globally homogenous, solution. The high-order interaction is the driver of the resulting patchy states, which emerge as follow a symmetry breaking instability caused by the injection of a  tiny non homogeneous perturbation. This can be though as a generalisation of the Turing instability on hypergraphs. In particular, we considered the interaction mediated by the number of interacting neighbouring units, namely the size of the hyperedge, and a diffusive-like process, again biased by the number of  neighbours. In both cases we provided sufficient conditions for the emergence of spatial patterns. 

Our findings have been corroborated by numerical simulations applied to two reference models. A Volterra model that describes the interaction among predators and prey in ecological niches, and the Brusselator model, a prototype model of nonlinear dynamics, that describes the interaction among reacting and diffusing chemicals.

The proposed framework goes beyond the examples hereby presented and, because of its generality,  it could prove useful in tackling those problems were simultaneous many-body interactions within a complex environment are to be properly accounted for.

\bibliographystyle{plain}
\bibliography{bib_HRW}
\end{document}